\documentclass[aps,prb,preprint,superscriptaddress]{revtex4-1}
\usepackage{graphicx}
\usepackage{subfigure}
\usepackage{color}
\usepackage{amsmath}
\usepackage[hang,flushmargin]{footmisc} 
\newcommand{\We}{W\hspace{-0.25em}e}
\newcommand{\Oh}{O\hspace{-0.1em}h}

\begin{document}

\preprint{}

\title{Symmetry-breaking in drop bouncing on curved surfaces}

\author{Yahua Liu$^{1\dagger \S}$, Matthew Andrew$^{2\dagger}$, Jing Li$^1$, Julia M Yeomans$^{2\ast}$, Zuankai Wang$^{1,3\ast}$
\renewcommand{\thefootnote}{\fnsymbol{footnote}}
\setcounter{footnote}{\ast}
\footnotetext[0]{$^\dagger$These authors contributed equally to this work. $^\ast$Correspondence should be addressed to Z.W (zuanwang@cityu.edu.hk) or J.M.Y. (j.yeomans1@physics.ox.ac.uk). $^\S$Present Address: Key Laboratory for Precision \& Non-traditional Machining Technology of Ministry of Education, Dalian University of Technology, Dalian 116024, China.}
}
\noaffiliation
\affiliation{Department of Mechanical and Biomedical Engineering, City University of Hong Kong, Hong Kong 999077, China}
\affiliation{The Rudolf Peierls Centre for Theoretical Physics, 1 Keble Road, Oxford, OX1 3NP, UK}
\affiliation{Shenzhen Research Institute of City University of Hong Kong, Shenzhen 518057, China}

\begin{abstract}
\normalsize
{\bf
The impact of liquid drops on solid surfaces is ubiquitous in nature, and of practical importance in many industrial processes. A drop hitting a flat surface retains a circular symmetry throughout the impact process; however a drop impinging on \textit{Echevaria} leaves exhibits asymmetric bouncing dynamics with distinct spreading and retraction along two perpendicular directions. This is a direct consequence of the cylindrical leaves which have a convex/concave architecture of size comparable to the drop. Systematic experimental investigations on mimetic surfaces and lattice Boltzmann simulations reveal that this novel phenomenon results from an asymmetric momentum and mass distribution that allows for preferential fluid pumping around the drop rim. The asymmetry of the bouncing leads to $\sim$ 40\% reduction in contact time. We expect that the coupling of fast drop detachment on surfaces of different architectures (convex, concave or corrugated) with facile and scalable manufacturing has potential in a wide range of applications.}
\end{abstract}

\maketitle

Since Worthington's pioneering work studying  the complex dynamics of liquid drops impacting on solid surfaces\cite{Worthington1876} in 1876, extensive progress has been made in understanding and controlling drop dynamics on various textured surfaces. Recent research in particular has shown that the spreading and retraction dynamics of impacting drops is highly dependent on both the roughness and the wettability of the underlying substrate\cite{Yarin2006, VaranasiDeng2009, VaranasiBird2013, QuereGauthier2015, BhushanJung2008, Bayeryeong2014, RubinsteinKolinski2014, MugeleJolet2015, LohseTsai2009, NagelXu2005, PetitjeansLagubeau2012, MugeledeRuiter2012, VovelleBergeron2000, WangHao2015}. Progress has been driven by the intrinsic scientific interest and beauty of fluid impacts\cite{TakeharaThoroddsen2008}, together with recent advances in the ability to fabricate micro- and nano-scale surfaces\cite{ButtDeng2012, RobinVerho2012}, and also because drop impact is central to many technological processes including DNA microarrays, digital lab-on-a-chip, water harvesting, dropwise heat removal and anti-icing\cite{AizenbergMishchenko2010, JiangJu2012, WangChen2011, ChenJonathan2009, Stone2012, PoulikakosMaitra2014, PoulikakosMaitra2014-1}.

Drops hitting superhydrophobic surfaces can bounce off quickly because of the low friction between drop and the substrate, either at the end of retraction\cite{QuereRichard2002, Marmur2004} or at their maximum extension in a pancake shape\cite{Zwliu2014, WangLiu2015, JuliaMoevius2014}. Normally the drop retains a circular symmetry during the bouncing and the contact time is bounded below by the Rayleigh limit\cite{Rayleigh1879}. However Bird {\it et al.}\cite{VaranasiBird2013} showed that drops impacting on surfaces where asymmetry is introduced with ridges an order of magnitude smaller than the drop leave the surface with shortened contact time. Moreover, the contact time of drops bouncing on a superhydrophobic macrotexture can take discrete values depending on the drop impact point relative to the texture and its impact velocity\cite{QuereGauthier2015}. The left-right symmetry can be also broken by imposing a surface gradient to induce a directional movement\cite{Wangchu2010, Chendaniel2001, Demirelmalvadkar2010, Demirelmalvadkar2014, WangMalouin2010} or by considering impacts on a moving surface\cite{StoneBird2009, WangChen2005}. In these studies, the reported surfaces are still macroscopically flat, with the feature size at the scale of microns or nanometres. Inspired by the observation that many natural surfaces have much larger-scale convex or concave architecture, in this work we first consider the impact of drops on a natural $Echevaria$ surface.

\noindent  \textbf{Results}\\
\noindent \textbf{Asymmetric bouncing on natural surfaces.} Figures 1a-c show the optical and scanning electron microscopic (SEM) images of the surface. The {\it Echevaria} leaves approximate cylinders with a diameter of curvature ($D$) a few millimeters. The surface of the leaves is covered by waxy nanofibers\cite{ChenZhang2015} yielding an apparent contact angle over $160^{\circ}$. Our experimental results are are very different from those conventionally reported on a flat superhydrophobic surface\cite{VaranasiDeng2009, BhushanJung2008}. Figure~1d presents selected snapshots of a drop of diameter $D_0=2.9$ mm impinging on the convex surface of an {\it Echeveria} leaf with a diameter of curvature  of $\sim$ 8.2 mm. The impact velocity is $0.63 \, \mathrm{m \, s^{-1}}$, corresponding to ${\We}=7.9$ and $\Oh$=0.0028. Here, $\We=\rho v_0^2 r_0 /\gamma$ is the Weber number, where $r_0$ is the drop radius, $\rho$ is the liquid density and $\gamma$ is the liquid-vapor surface tension, and  $\Oh=\mu/\sqrt{\rho \gamma r_0}$ is the Ohnesorge number, with $\mu$ the liquid viscosity. The impacting drop initially spreads isotropically, but the drop spreading becomes increasingly anisotropic as  the drop starts to retract (Supplementary Movie 1). Interestingly, when the liquid in the axial (straight) direction has started to retract at $\sim$ 3.8 ms, the liquid in the azimuthal (curved) direction continues to spread, with a sustained fluid transfer from the axial direction. Once the fluid in the axial direction has fully contracted at 11.8 \,ms ($=1.73\sqrt{\rho r_0^3/\gamma}$), the drop leaves the surface maintaining an elongated shape along the azimuthal direction, indicating that the asymmetric bouncing can be driven by preferential retraction along just one axis. The contact time ($t_0$) is $\sim 30\%$  faster than that on the equivalent flat substrate\cite{QuereRichard2002, RubinsteinKolinski2014, MugeleJolet2015} (which is compared in Supplementary Movie 2, right and Supplementary Fig.~1) and sphere (Supplementary Fig.~2). A reduction in contact time due to bouncing asymmetry was first reported by Bird {\it et al.}\cite{VaranasiBird2013} who considered drop breakup on surfaces with ridges of size $\sim$100 microns.

\noindent \textbf{Symmetry-breaking in droplet bouncing on synthetic surfaces.} Inspired by this unexpected result we hypothesize that new physics comes into play when symmetry-breaking mechanisms are introduced by the convexity of the surface\cite{Vignes-AdlerLorenceau2009, VandewalleGilet2010}. To explore these, we fabricated curved surfaces with varying diameters of curvature $D$ between 4 mm and 20 mm (Supplementary Fig.~3). The surfaces are coated with hydrophobic rosettes of diameter $\sim5$ $\mu$m to give an intrinsic contact angle of $\sim160^{\circ}$.

Drop impact on the fabricated surfaces reveals similar bouncing dynamics to that on the natural surface. Figure~2a shows the time-evolution of the spreading diameters in the axial and azimuthal directions on the curved surface with $D=8$ mm at $\We=7.9$ (Supplementary Movie 2, left). The maximal spreading diameter in the axial direction is measured to be 4.45 mm, following the scaling\cite{QuereClanet2004} of $\sim \We^\frac{1}{4}$. To quantify the spreading asymmetry we define $k$ as the ratio of the maximum values of the drop spreading diameters in the azimuthal and axial directions. Figure~2b plots the variation of $k$ as a function of the diameter of curvature normalized by the initial drop diameter ($D/D_0$). It is apparent that an increase in the structural anisotropy gives rise to a larger $k$. Note that for the flat or spherical superhydrophobic surfaces, the $k$ is equivalent to unity, suggesting that the asymmetric spreading is modulated by the structural anisotropy. Figure~2c plots the variation of the contact time as a function of normalized diameter of curvature for different $\We$. The contact time is also significantly affected by the anisotropy: at a constant $\We$, the symmetry-breaking surfaces with smaller diameters of curvature corresponds to smaller contact times. To better elucidate the dependence of the contact time on the surface structure, we decompose the contact time into the spreading time $t_1$ and retraction time $t_2$ along the axial direction in Fig.~2d. It is apparent that the spreading time is almost independent of surface curvature, partially due to the fact that the spreading is mainly dominated by the inertia. However the retraction time shows a strong decrease with decreasing $D$: for $D=6$ mm the total decrease in contact time compared to a flat substrate is $\sim 40\%$ for $\We \sim 15$. These results, in conjunction with the spreading dynamics shown in Fig.~2a and b, indicate that the asymmetric bouncing is indeed modulated by the asymmetric curvature whose size is comparable to that of the impacting drop. This argument is also confirmed by our control experiment on the spherical surface where the bouncing is symmetric and the contact time is the same as that on a flat surface (Supplementary Fig.~2).

\noindent \textbf{Mechanisms for symmetry-breaking and contact time reduction.} To interpret the mechanism behind the asymmetric bouncing observed in our experiments, we first considered the effect of surface topography on initial drop momentum. Two factors will lead to more momentum being transferred in the azimuthal direction than in the axial direction. First, distinct from the flat surface, the impact area on the curved surface is approximately elliptical. As a result of such an asymmetric footprint, more momentum will be transferred perpendicular to the long axis of the ellipse, i.e., along the azimuthal direction. Second, fluid landing on the curved sides of the asymmetric surface has a tangential component of momentum which will continue unperturbed.

To test this interpretation, we modelled the drop impact, using a lattice Boltzmann algorithm to solve the continuum equations of motion of the drop\cite{Lintaehun2010, TaehunKevin2013}. Details of the equations and the numerical algorithm are given in the Supplementary Information. Figure~3a shows snapshots of the time evolution of a drop impacting on the asymmetric surface obtained from the numerics. Comparison of the simulation results with those obtained in the experiments (Fig. 1d and Supplementary Movie 1) shows that the evolution of the drop shape is qualitatively the same during the rebound, with the axial direction starting to retract first. The colour shading represents the relative heights of the fluid at each time, red high to blue low. Note that in particular (the 3$^{rd}$ image) a large rim develops in the azimuthal direction. The arrows in the figure show the local fluid velocity field. Consistent with our experimental observation, during the initial spreading stage, the fluid exhibits a radial outwards flow. As spreading progresses, there is a preferential flow to the azimuthal direction that drives the formation of the larger liquid rims. The liquid pumped around the rim to the azimuthal direction acts to amplify the contrast in the drop retraction between the two directions, leading to a positive feedback that enhances the asymmetry of the bouncing. This scenario is in striking contrast to that on the flat surface.

The simulations allow us to understand how the asymmetric surface topography affects the drop bouncing. Figure~3b displays the variation of the momentum in the horizontal direction relative to the initial impact momentum as a function of time during the impact on the surface with $D/D_0=1.2$. In the figure, a positive momentum corresponds to drop spreading whilst a negative momentum corresponds to drop retraction. From the graph, it can be clearly seen that the momentum in the azimuthal direction is always larger than that in the axial direction. When the momentum in the axial direction starts to reverse its direction, the azimuthal momentum remains positive and indeed increases slightly. This is consistent with our experimental observations that the drop sustains a spreading state without retracting in the azimuthal direction, demonstrating the positive feedback from the axial direction that enables this to occur. As a comparison, we also plotted the variation of momentum on a flat surface (blue curve in Fig.~3b), which shows that the momentum along any given direction for a flat substrate lies between the two curves for the asymmetric surface. In order to quantify how the momentum anisotropy is dictated by the surface topography, we calculated the ratio between the maximal momentum in the azimuthal direction and that in the axial direction. As shown in Fig.~3c, the momentum anisotropy decreases with increasing diameter of curvature.

To further validate that the momentum asymmetry is responsible for the asymmetric bouncing, we simulated drop impact on a flat surface by introducing a momentum asymmetry into the simulation manually immediately after the initial collision. The momentum in the azimuthal direction was increased by a factor of 2 while the momentum in the axial direction was reduced by a factor of 2. Indeed, as shown in Supplementary Fig.~4, the drop shows qualitatively the same bouncing pathway as that in Fig. 3a. In particular, the drop retraction in the axial direction is much faster than that in the azimuthal direction, which is consistent with our experimental results. Moreover, to validate that a momentum asymmetry can be induced by an elliptical drop footprint, we also simulated an initially elliptical drop impacting a flat substrate which, again, led to a very similar asymmetric bouncing (Supplementary Fig.~5).

For surface obstacles much smaller than the drop, such as those used in Bird {\it et al.}\cite{VaranasiBird2013}, the initial momentum asymmetry is largely suppressed. Notably, the plot of the variation of the contact time relative to that on a flat surface as a function of $D/D_0$ (Fig. 3d) displays a minimum at $D/D_0 \sim 1$. This is expected since, as the obstacle size decreases the surface becomes more comparable to a flat surface ($D/D_0 \sim 0$) and the momentum anisotropy starts to gets smaller and has less effect. However, in this regime, the critical Weber number for drop splitting ($We_c$) is low (Fig. 3e) and drops in the experiments tend to break up giving the mechanism for contact time reduction described by Bird {\it et al.} The drop retracts faster along the ridge than perpendicular to it. As a result it tends to fragment and the newly formed inner rims retract away from the obstacle resulting in the contact time reduction. These two regimes for contact time reduction serve to emphasise the richness of the physics underlying bouncing on curved and irregular surfaces.

We performed a simple hydrodynamic analysis to explain the contact time reduction associated with the asymmetric bouncing. Since the drop spreading is mainly governed by the inertia, we consider the drop retraction process here. The drop retraction is primarily driven by the decrease in surface energy of the thinner central film which leads to a force pulling the rim of the drop inwards. For conventional bouncing, the drop retraction is symmetric and the surface energy of a central film of radius $r$ is $E_s\approx \pi r^2 \gamma (1-\cos \theta)$, where $\theta$ is the apparent contact angle, giving a retraction force\cite{Bonnbartolo2005}, $F_s=\frac{\partial E_s}{\partial r} \approx 2 \pi r \gamma(1-\cos \theta)$. Fast drop detachment requires not only a large driving force in the central film, but also a small inertia of the rim. However, due to the symmetric retraction and mass conservation, these two processes are mutually exclusive, since a reduction in the central film radius $r$ leads to an increase in the mass of the liquid rim  as is apparent in Supplementary Fig.~1.

Interestingly, for the symmetry-breaking surface this conflict is resolved by the preferential fluid flows around the drop rim (Figs. 3a and 4a). To demonstrate this, in Fig.~4c we showed selected plan-view images of drop retraction on the asymmetric surface. The panels just above (Fig. 4b) and below (Fig. 4d) this figure correspond to the side view of drop retraction in the axial direction and azimuthal direction, respectively. It is clear that owing to the preferential liquid pumping to the azimuthal direction (blue arrows), the size of rim in the axial direction is almost unchanged throughout the retraction stage, whereas that in the azimuthal direction shows a significant increase. This is confirmed by the simulation results in Fig.~4e, which plots the time evolution of the rim heights in the two directions and the height of the central film. Notably, the mass per unit length of the rim along the azimuthal direction is more than twice that in the axial direction when the rims from opposite sides of the drop meet just prior to bouncing, confirming the apparent symmetry-breaking in the mass distribution. Moreover, the desirous reduction in the mass of rim in the axial direction is achieved without compromising the retraction force. Due to the preferential spreading on the curved surface, the central film can be approximated by an ellipse with a major axis $b$ (in the azimuthal direction) and minor axis $a$ (in the axial direction). Thus, the surface energy of the central film is $E_a \approx \pi a b \gamma (1-\cos \theta)$. As the asymmetric retraction proceeds the length of the major axis $b$ remains constant while there is a continuous reduction in $a$. Hence the retraction force is now $F_a \approx \pi b \gamma (1-\cos \theta)$, and the ratio of the force acting on the rim on the curved surface to that on the flat surface is $b/2r$. Experimentally, the drop diameter $2r$ on the symmetric surface continually decreases whilst the azimuthal diameter  $b$ on the curved surface remains unchanged. Thus, the synergy of the enhanced retraction force and reduced mass of rim rendered by the symmetry-breaking structure results in a remarkably efficient pathway for fast drop retraction. The convergence of liquid in the axial direction translates into motion perpendicular to the surface and drives the drop upwards. Moreover, as shown in Fig. 4d, as the drop retracts on the curved surface, the surface tension energy converts to kinetic energy with a velocity component in the vertical direction (red arrows) which will aid the bouncing. By contrast, on the flat surface, the drop transition from the oblate shape (7.6 ms) to a prolate one (16.1 ms) prior to its jumping takes a longer time (Supplementary Fig. 1).

\noindent  \textbf{Discussion}\\
In a broad perspective, we expect that rapid bouncing driven by an asymmetric momentum transfer occurs on many surfaces that have asymmetric structure on the order of the drop size. The most obvious extension is to a surface which is concave in one direction and flat in the perpendicular direction. Fig.~5 shows snapshots of a water drop hitting a concave surface with a diameter of curvature $D=-8$ mm at $We=7.9$. By contrast to the impact on the convex surface, there is a preferential fluid flow from the azimuthal direction to the axial direction. After the drop reaches its maximum spreading in the azimuthal direction at $\sim 3.0$ ms, the liquid in the axial direction continues to spread and finally leaves the surface in a shape elongated along the axial direction at $\sim 10.3$ \,ms ($=1.58\sqrt{\rho r_0^3/\gamma}$) (Supplementary Movie 3). Indeed the contact time reduction is even more pronounced than that on the convex surface, suggesting that corrugated surfaces may be excellent candidates for enhanced water repellency and other applications. For example, many pathogens and diseases are transmitted through drops\cite{Savage2015, BourouibaGilet2014}, and thus the fast drop detachment from natural plant and our synthetic surfaces might significantly decrease the likelihood of virus and bacteria deposition. Additionally, the presence of large-scale curved topography on corrugated surfaces could offer promise for enhanced heat transfer performances and anti-icing\cite{ChenJonathan2009, AizenbergMishchenko2010, Stone2012}. Moreover, these corrugated surfaces with such millimeter-scale features are scalable in manufacturing. Thus we envision that the asymmetric bouncing discovered on curved surfaces not only extends our fundamental understanding of classical wetting phenomenon, but also offers potential for a wide range of applications\cite{Savage2015, BourouibaGilet2014, PoulikakosMaitra2014, WangChen2011, AizenbergMishchenko2010}.
\vspace{12pt}

\noindent  \textbf{Methods}

\noindent \textbf{Preparation of the asymmetric surfaces.} The asymmetric convex surfaces were fabricated on copper plate by combined mechanical wire-cutting and chemical etching. A series of convex surfaces were first cut with arc diameters ranging from $6 \, \mathrm{mm}$ to $20 \, \mathrm{mm}$. Then the as-fabricated surfaces were coated with hydrophobic rosettes of average diameter $\sim 5.0 \, \mathrm{\mu m}$ to render them superhydrophobic. More specifically, after ultrasonic cleaning in ethanol and deionized water for $10 \, \mathrm{min}$, respectively, the surfaces were washed by diluted hydrochloric acid (1 M) and deionized water, followed by drying  in nitrogen stream. They were then immersed in a freshly mixed aqueous solution of $2.5 \, \mathrm{mol \, l^{-1}}$ sodium hydroxide and $0.1 \, \mathrm{mol \, l^{-1}}$ ammonium persulphate at room temperature for $\sim 60 \, \mathrm{min}$, after which they were fully rinsed with deionized water and dried again in nitrogen stream. After the oxidation process, the surfaces were uniformly coated by $\mathrm{CuO}$ flowers of diameter $\sim 5.0 \, \mathrm{\mu m}$. All the surfaces were modified by silanization by immersion in a 1 mM n-hexane solution of trichloro-(1H,1H,2H,2H)-perfluorooctylsilane for $\sim 60 \, \mathrm{min}$, followed by heat treatment at $\sim 150 \, \mathrm{{}^\circ C}$ in air for 1 h to render superhydrophobic.

\vspace{12pt}

\noindent \textbf{Characterization of the surfaces}. The optical images of $Echeveria$ were recorded by a Nikon digital camera (Digital SLR Camera D5200 equipped with a Micro-Nikkor 105 mm f/2.8G lens). The micro/nano structures of protuberances and nanofibers were characterized by a field-emission scanning electron microscope (Quanta$^{TM}$ 250 FEG).  Due to the extremely low conductivity of the $Echeveria$ surface, a thin layer of carbon was coated on the surface before the SEM measurement.

\vspace{12pt}

\noindent \textbf{Contact angle measurements.} Owing to the asymmetric structure, it is difficult to measure the contact angle of the as-fabricated asymmetric surfaces. For the flat surface coated with rosettes (subject to the same treatment), the apparent, advancing ($\theta_a$) and receding contact angles ($\theta_r$) are $163.4^{\circ} \pm 2.6^{\circ}$, $165.1^{\circ}\pm 2.1^{\circ}$ and $161.9^{\circ} \pm 1.5^{\circ}$, respectively. These values are the average of five measurements.

\vspace{12pt}

\noindent \textbf{Impact experiments.} Impact experiments were performed in ambient environment, at room temperature with 60\% relative humidity. Briefly, the Milli-Q water drop of $\sim 13 \, \mathrm{\mu L}$ (with drop diameter $\sim 2.9 \, \mathrm{mm}$) was released from a fine needle equipped with a syringe pump (KD Scientific Inc.) at different heights to vary the impact velocity upon the substrate. The impact dynamics of drop was captured simultaneously from side view and plan view by using two synchronous high speed cameras (Fastcam SA4, Photron limited) at a frame rate of 10,000 fps with a shutter speed 1/93,000 sec. The configuration of the drops during impact was measured using ImageJ software (Version 1.46, National Institutes of Health, Bethesda, MD).

\makeatletter
\def\bib@device#1#2{}
\makeatother

\vspace{12pt}

\noindent \textbf{Acknowledgments}\\
We acknowledge support from the Hong Kong General Research Fund (No. 11213414),
National Natural Science Foundation of China (No. 51475401) to Z.W., ERC Advanced Grant, MiCE, to J.M.Y.

\vspace{12pt}

\noindent \textbf{Author contributions}\\
Y.L. and Z.W. conceived the research. Z.W. and J.M.Y. supervised the research. Y.L. and J.L. designed and carried out the experiments. Y.L. and M.A. analysed the data. M.A. and J.M.Y. ran the simulations. Z.W., J.M.Y., Y.L. and M.A. wrote the manuscript. Y.L. and M.A. contributed equally to this work.

\vspace{12pt}

\noindent \textbf{Additional information}\\
Supplementary information is available in the online version of the paper. Reprints and permissions information is available online at www.nature.com/reprints. Correspondence and requests for materials should be addressed to Z.W. or J.M.Y.

\vspace{12pt}

\noindent \textbf{Competing financial interests}\\
The authors declare no competing financial interests.

\begin{figure*}[htbp]
  \centering
  \includegraphics[clip=true, viewport=0 0 1000 600, keepaspectratio, width=1.05\textwidth]{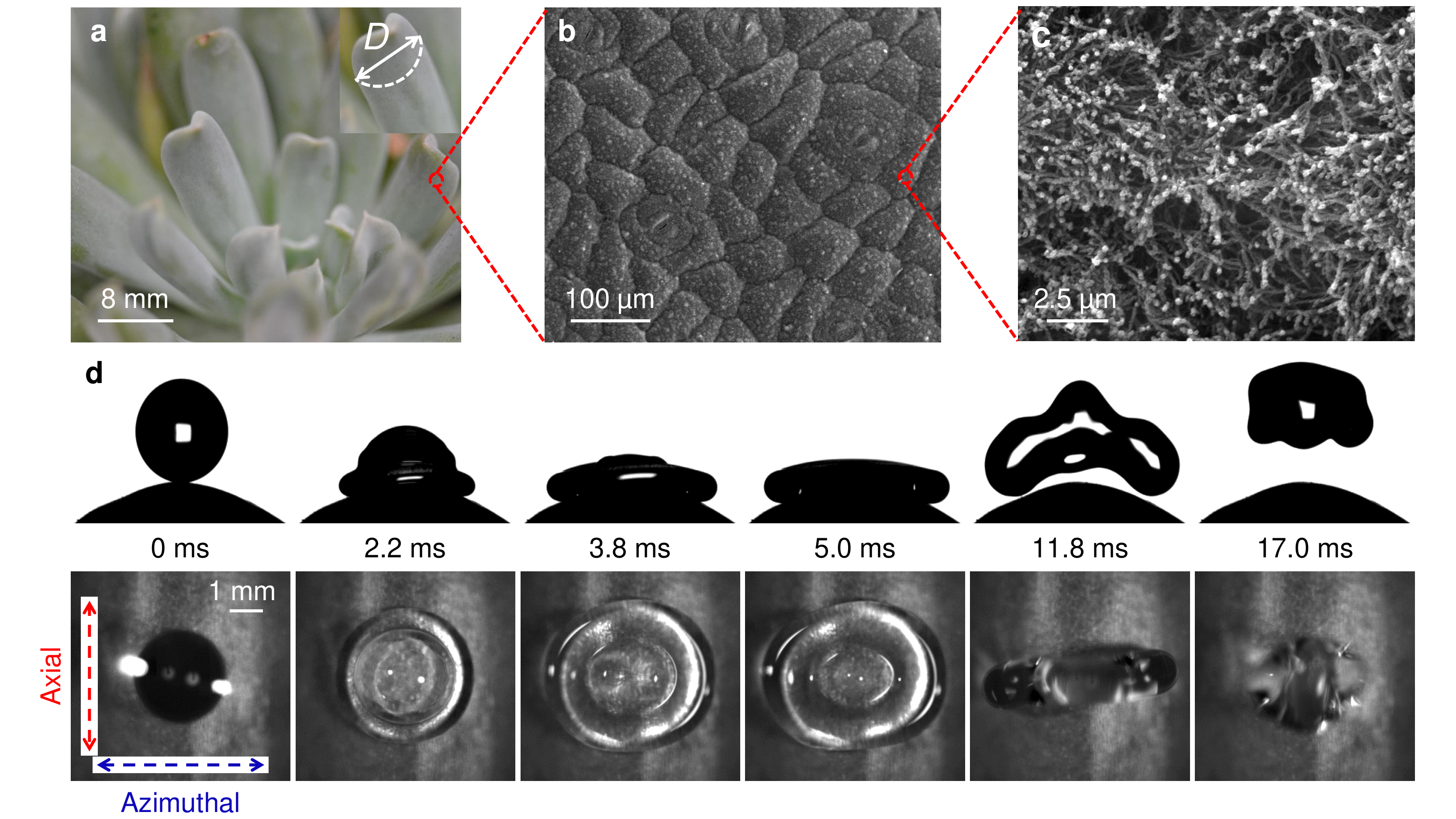} \\
  \renewcommand{\figurename}{\textbf{Figure}}
  \renewcommand{\thefigure}{\textbf{1 $|$}} 
  \caption {\linespread{1.5}\selectfont{\textbf{The surface morphology of \textit{Echeveria} and drop impact dynamics.} \textbf{a}, Optical image of {\it Echeveria} showing the curvature of individual leaves. \textbf{b}, Low-resolution scanning electron microscope (SEM) image of an {\it Echeveria} surface showing protuberances on the scale of 100 microns. \textbf{c}, Magnified SEM image of a single protuberance consisting of countless nanofibers. \textbf{d}, Selected snapshots showing a drop ($D_0=2.9$ mm) impacting on an {\it Echeveria} leaf at $\We=7.9$. The first row is a cross section parallel to the azimuthal direction and the second is a plan view from above the drop. After spreading to its maximum extension in the axial direction at 3.8 ms, the drop continues its spreading in the azimuthal direction and reaches its maximum spreading at 5.0 ms, while retracting in the axial direction. The drop bounces off the surface at 11.8 ms with a much shortened contact time compared to that on the flat surface (16.1 ms, Supplementary Fig.~1).}}

  \label{fig:fig1}
\end{figure*}

\begin{figure*}[htbp]
  \centering
  \includegraphics[clip=true, viewport=0 0 980 800, keepaspectratio, width=\textwidth]{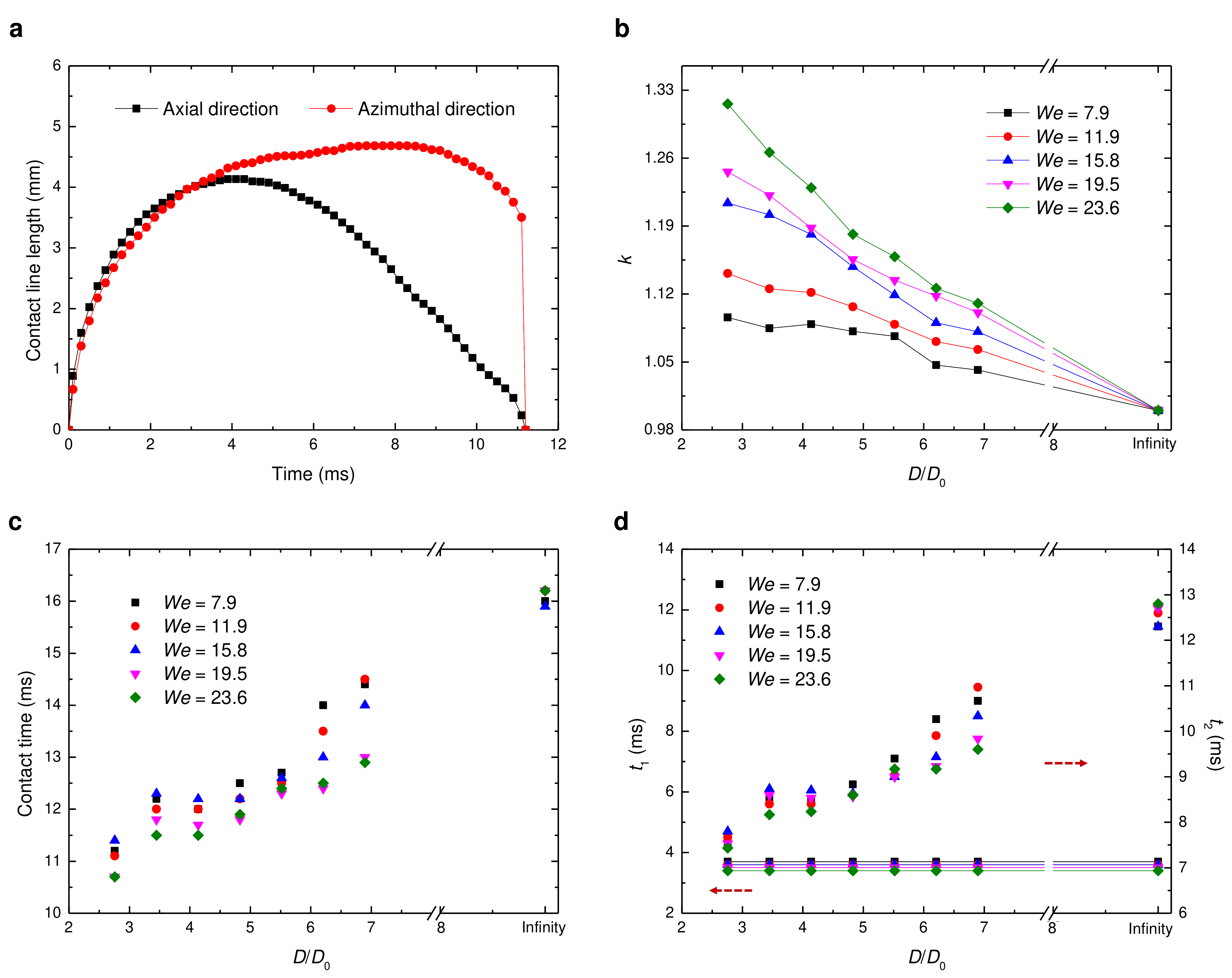} \\
  \renewcommand{\figurename}{\textbf{Figure}}
  \renewcommand{\thefigure}{\textbf{2 $|$}} 
  \caption {\linespread{1.5}\selectfont{\textbf{Asymmetric bouncing on bioinspired asymmetric surface.} \textbf{a}, The variations of contact line length in the axial and azimuthal directions as a function of time. The drop continues spreading in the azimuthal direction after it reaches its maximum extension in the axial direction at 3.8 ms. While retracting in the axial direction, the lateral extension in the azimuthal direction remains almost constant. \textbf{b}, \textbf{c} and \textbf{d}, The variation of the $k$ (defined as the ratio of the maximum spreading diameters in the azimuthal and axial directions), the contact time, the axial spreading time $t_1$ (left) and axial retraction time $t_2$ (right) as a function of surface curvature $D$ (normalised by drop radius $D_0$) under different $\We$. }}

  \label{fig:fig2}
\end{figure*}

\begin{figure*}[htbp]
  \centering
  \includegraphics[clip=true, viewport=0 0 990 1000, keepaspectratio, width=\textwidth]{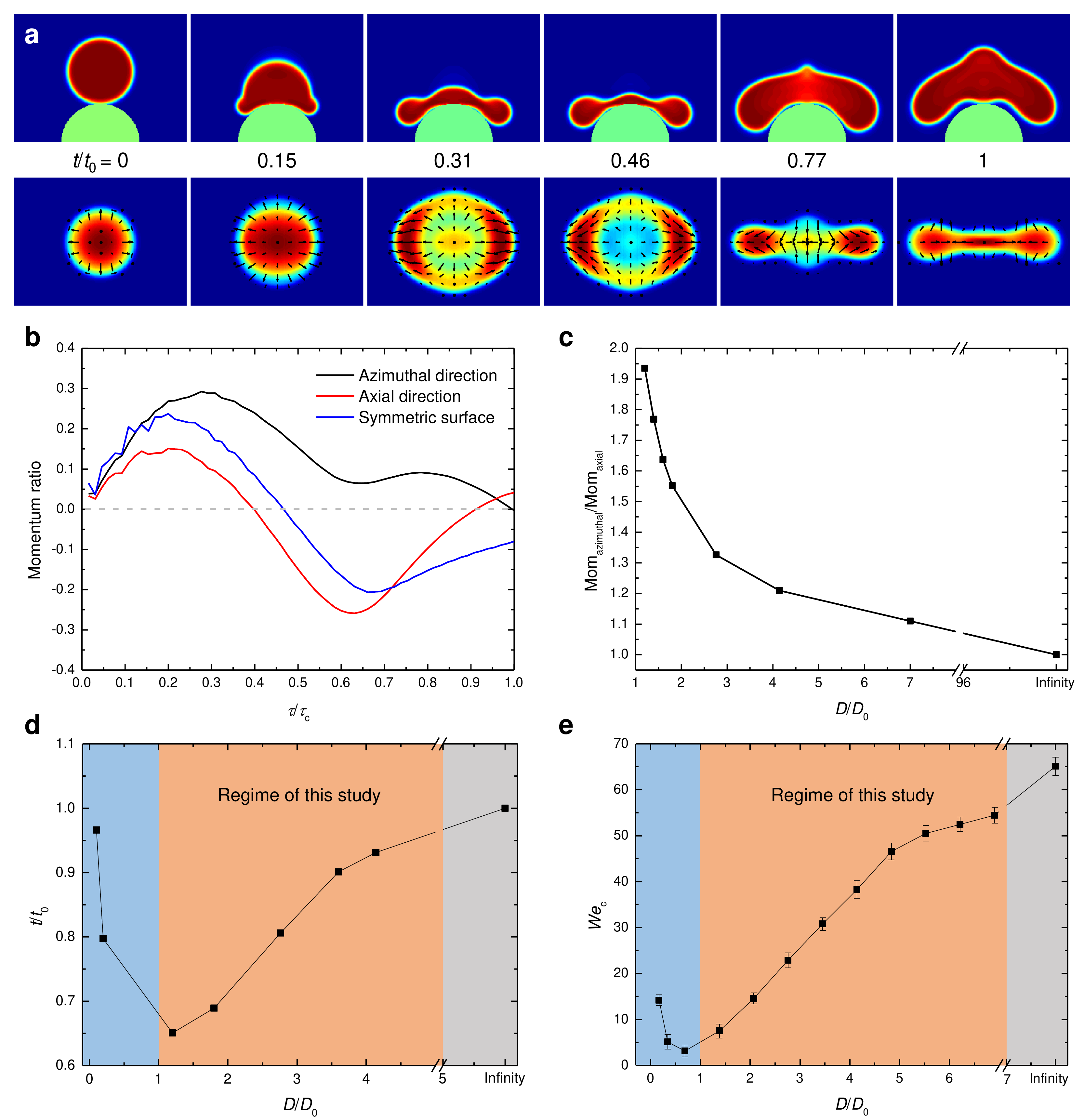} \\
  \renewcommand{\figurename}{\textbf{Figure}}
  \renewcommand{\thefigure}{\textbf{3 $|$}} 
  \caption {\linespread{1.5}\selectfont{\textbf{Asymmetric bouncing verified by simulation.} \textbf{a}, Selected snapshots obtained using the lattice Boltzmann simulation showing the time evolution of a drop bouncing on an asymmetric surface for $\We = $10.6 and $\Oh =$ 0.0028. The top panel corresponds to the cross-section view parallel to the azimuthal direction, and the bottom panel is the plan view from above the drop. The colours in the plan view are indicative of the relative height of the liquid at each time (see Fig. 4e for quantitative data) and arrows indicate the velocity flux. \textbf{b}, The time evolution of the momentum normalised by the total initial momentum along the axial (red) and azimuthal (black) directions on the curved surface ($D/D_0 = $1.2), and for any direction on a flat surface (blue),}}
  \label{fig:fig3}
\end{figure*}

\clearpage

\noindent under the same $\We = $10.6 and $\Oh =$ 0.0028. The positive values correspond to the spreading stage whilst the negative values correspond to drop retraction. \textbf{c}, The ratio between the maximum momentum transferred into the azimuthal direction and the axial direction as a function of normalized surface curvature $D/D_0$. \textbf{d}, Simulations of the variation of the contact time relative to that for a flat superhydrophobic surface showing a different dependence on $D/D_0$ in different regimes. \textbf{e}, Experimental results for the variation of the critical $We$ for drop break-up ($We_c$) with $D/D_0$. Error bars denote the range of the measurements.

\begin{figure*}[htbp]
  \centering
  \includegraphics[clip=true, viewport=0 0 1050 1000, keepaspectratio, width=\textwidth]{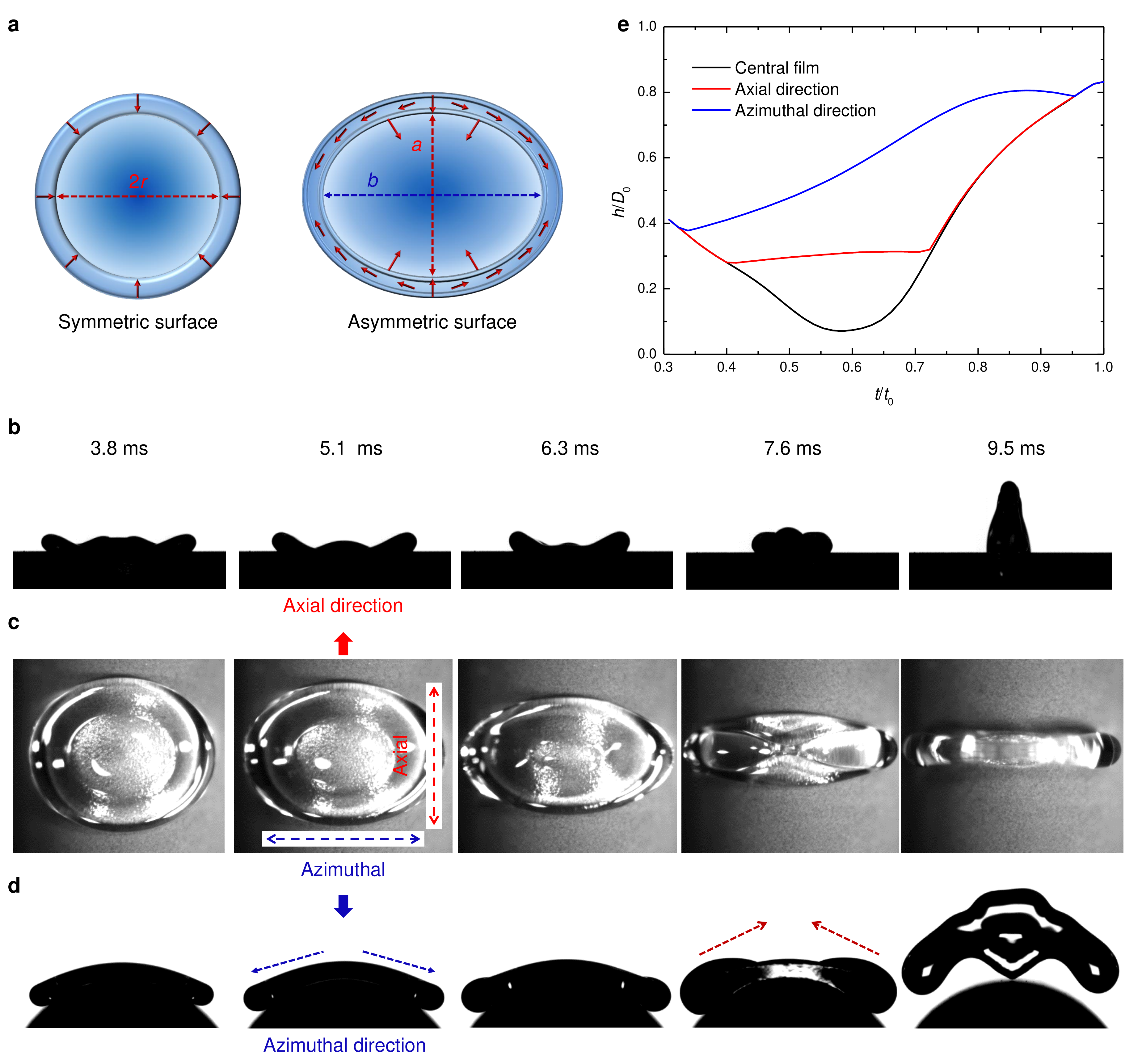}\\
  \renewcommand{\figurename}{\textbf{Figure}}
  \renewcommand{\thefigure}{\textbf{4 $|$}} 
  \caption {\linespread{1.5}\selectfont{\textbf{Drop retraction and bouncing dynamics.} \textbf{a}, Schematic drawings of the rim and central film on the symmetric and asymmetric surfaces, respectively. On the symmetric surface, the rim retracts uniformly inwards towards the central film. On the asymmetric surface, the central film is an ellipse with major axis {\it b} (in the azimuthal direction) and minor axis {\it a} (in the axial direction) and retraction is primarily along the axial direction. \textbf{b}, Selected side-view images of drop retracting on the asymmetric surface in the axial direction. The drop rim is drawn inwards by the central film and the size of rim remains almost unchanged in the majority of retraction process. \textbf{c}, Selected plan-view images of drop retracting on the asymmetric surface. \textbf{d}, Selected side-view images of drop retracting in the azimuthal direction. Due to preferential liquid pumping}}
  \label{fig:fig4abcde}
\end{figure*}

\clearpage

\noindent around the rim (blue arrows), there is a continuous increase in the height of the azimuthal rim. Moreover, as the drop retracts on the curved surface, the surface tension energy converts to kinetic energy with a velocity component in the vertical direction (red arrows). \textbf{e}, Comparison of the time evolution of the normalized rim heights in the axial and azimuthal directions based on the simulation. The height is scaled by the drop diameter $D_0$ and time by the contact time $t_0$. During the retraction stage the axial rim height stays roughly constant whilst the azimuthal rim height increases greatly due to the preferential flow and mass transfer. The reduced mass of the axial rim rendered by the symmetry-breaking flows results in a remarkably efficient pathway for fast drop retraction.

\begin{figure*}[htbp]
  \centering
  \includegraphics[clip=true, viewport=0 0 900 420, keepaspectratio, width=\textwidth]{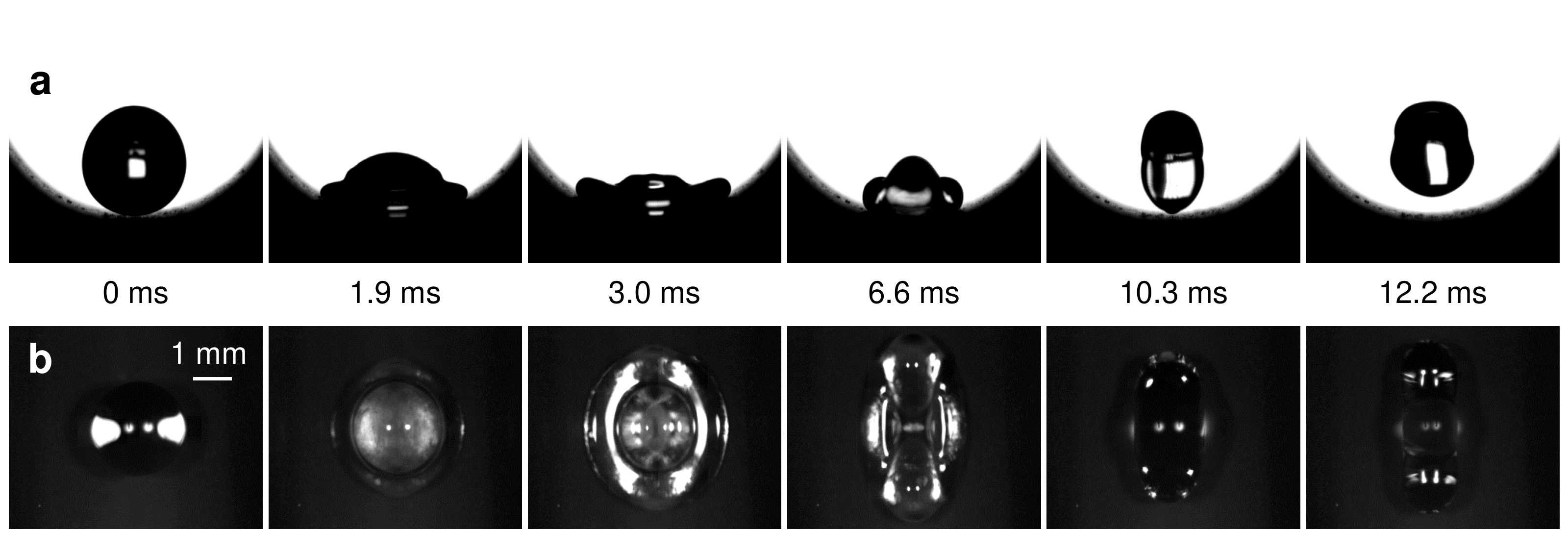}\\
  \renewcommand{\figurename}{\textbf{Figure }}
  \renewcommand{\thefigure}{\textbf{5 $|$}} 
  \caption {\linespread{1.5}\selectfont{Selected snapshots showing a drop ($D_0=2.9 \, \mathrm{mm}$) impacting on the concave surface (with a diameter of curvature $D=8.0$ mm) at $\We$ = 7.9 both from the side view (\textbf{a}) and plan view (\textbf{b}). After spreading to its maximum extension in the azimuthal  direction at  3.0 ms, the drop continues its spreading in the axial direction and reaches its maximum spreading at 6.6 ms. It finally bounces off the surface after a contact time of 10.3 ms, reduced by $\sim$ 40\% compared to that on a symmetric surface. More details are shown in Supplementary Movie 3.}}
  \label{Fig:Fig5}
\end{figure*}


\vspace{12pt}

\noindent \textbf{References}
\begin{thebibliography}{10}
\expandafter\ifx\csname url\endcsname\relax
  \def\url#1{\texttt{#1}}\fi
\makeatletter
\def\@biblabel#1{#1.}
\setlength{\labelwidth}{2em}
\makeatother
\expandafter\ifx\csname urlprefix\endcsname\relax\def\urlprefix{URL }\fi
\providecommand{\bibinfo}[2]{#2}
\providecommand{\eprint}[2][]{\url{#2}}

\bibitem{Worthington1876}
\bibinfo{author}{Worthington, A.~M.}
\newblock \bibinfo{title}{On the forms assumed by drops of liquids falling
  vertically on a horizontal plate}.
\newblock \emph{\bibinfo{journal}{Proc. Roy. Soc. Lond.}}
  \textbf{\bibinfo{volume}{25}}, \bibinfo{pages}{261--272}
  (\bibinfo{year}{1876}).

\bibitem{Yarin2006}
\bibinfo{author}{Yarin, A.~L.}
\newblock \bibinfo{title}{Drop impact dynamics: Splashing, spreading, receding,
  bouncing ...}
\newblock \emph{\bibinfo{journal}{Annu. Rev. Fluid Mech.}}
  \textbf{\bibinfo{volume}{38}}, \bibinfo{pages}{159--192}
  (\bibinfo{year}{2006}).

\bibitem{VaranasiDeng2009}
\bibinfo{author}{Deng, T.} \emph{et~al.}
\newblock \bibinfo{title}{Nonwetting of impinging droplets on textured
  surfaces}.
\newblock \emph{\bibinfo{journal}{Appl. Phys. Lett.}}
  \textbf{\bibinfo{volume}{94}}, \bibinfo{pages}{133109}
  (\bibinfo{year}{2009}).

\bibitem{VaranasiBird2013}
\bibinfo{author}{Bird, J.~C.}, \bibinfo{author}{Dhiman, R.},
  \bibinfo{author}{Kwon, H.-M.} \& \bibinfo{author}{Varanasi, K.~K.}
\newblock \bibinfo{title}{Reducing the contact time of a bouncing drop}.
\newblock \emph{\bibinfo{journal}{Nature}} \textbf{\bibinfo{volume}{503}},
  \bibinfo{pages}{385--388} (\bibinfo{year}{2013}).

\bibitem{QuereGauthier2015}
\bibinfo{author}{Gauthier, A.}, \bibinfo{author}{Symon, S.},
  \bibinfo{author}{Clanet, C.} \& \bibinfo{author}{Qu{\'e}r{\'e}, D.}
\newblock \bibinfo{title}{Water impacting on superhydrophobic macrotextures}.
\newblock \emph{\bibinfo{journal}{Nat. Commun.}} \textbf{\bibinfo{volume}{6}},
  \bibinfo{pages}{8001} (\bibinfo{year}{2015}).

\bibitem{BhushanJung2008}
\bibinfo{author}{Jung, Y.~C.} \& \bibinfo{author}{Bhushan, B.}
\newblock \bibinfo{title}{Dynamic effects of bouncing water droplets on
  superhydrophobic surfaces}.
\newblock \emph{\bibinfo{journal}{Langmuir}} \textbf{\bibinfo{volume}{24}},
  \bibinfo{pages}{6262--6269} (\bibinfo{year}{2008}).

\bibitem{Bayeryeong2014}
\bibinfo{author}{Yeong, Y.~H.}, \bibinfo{author}{Burton, J.},
  \bibinfo{author}{Loth, E.} \& \bibinfo{author}{Bayer, I.~S.}
\newblock \bibinfo{title}{Drop impact and rebound dynamics on an inclined
  superhydrophobic surface}.
\newblock \emph{\bibinfo{journal}{Langmuir}} \textbf{\bibinfo{volume}{30}},
  \bibinfo{pages}{12027--12038} (\bibinfo{year}{2014}).

\bibitem{RubinsteinKolinski2014}
\bibinfo{author}{Kolinski, J.~M.}, \bibinfo{author}{Mahadevan, L.} \&
  \bibinfo{author}{Rubinstein, S.~M.}
\newblock \bibinfo{title}{Drops can bounce from perfectly hydrophilic
  surfaces}.
\newblock \emph{\bibinfo{journal}{Europhys. Lett}}
  \textbf{\bibinfo{volume}{108}}, \bibinfo{pages}{24001}
  (\bibinfo{year}{2014}).

\bibitem{MugeleJolet2015}
\bibinfo{author}{de~Ruiter, J.}, \bibinfo{author}{Lagraauw, R.},
  \bibinfo{author}{van~den Ende, D.} \& \bibinfo{author}{Mugele, F.}
\newblock \bibinfo{title}{Wettability-dependent bouncing on flat surfaces
  mediated by thin air films}.
\newblock \emph{\bibinfo{journal}{Nat. Phys.}} \textbf{\bibinfo{volume}{11}},
  \bibinfo{pages}{48--53} (\bibinfo{year}{2015}).

\bibitem{LohseTsai2009}
\bibinfo{author}{Tsai, P.~C.}, \bibinfo{author}{Pacheco, S.},
  \bibinfo{author}{Pirat, C.}, \bibinfo{author}{Lefferts, L.} \&
  \bibinfo{author}{Lohse, D.}
\newblock \bibinfo{title}{Drop impact upon micro- and nanostructured
  superhydrophobic surfaces}.
\newblock \emph{\bibinfo{journal}{Langmuir}} \textbf{\bibinfo{volume}{25}},
  \bibinfo{pages}{12293--12298} (\bibinfo{year}{2009}).

\bibitem{NagelXu2005}
\bibinfo{author}{Xu, L.}, \bibinfo{author}{Zhang, W.~W.} \&
  \bibinfo{author}{Nagel, S.~R.}
\newblock \bibinfo{title}{Drop splashing on a dry smooth surface}.
\newblock \emph{\bibinfo{journal}{Phys. Rev. Lett.}}
  \textbf{\bibinfo{volume}{94}}, \bibinfo{pages}{184505}
  (\bibinfo{year}{2005}).

\bibitem{PetitjeansLagubeau2012}
\bibinfo{author}{Lagubeau, G.} \emph{et~al.}
\newblock \bibinfo{title}{Spreading dynamics of drop impacts}.
\newblock \emph{\bibinfo{journal}{J. Fluid Mech.}}
  \textbf{\bibinfo{volume}{713}}, \bibinfo{pages}{50--60}
  (\bibinfo{year}{2012}).

\bibitem{MugeledeRuiter2012}
\bibinfo{author}{de~Ruiter, J.}, \bibinfo{author}{Oh, J.~M.},
  \bibinfo{author}{van~den Ende, D.} \& \bibinfo{author}{Mugele, F.}
\newblock \bibinfo{title}{Dynamics of collapse of air films in drop impact}.
\newblock \emph{\bibinfo{journal}{Phys. Rev. Lett.}}
  \textbf{\bibinfo{volume}{108}}, \bibinfo{pages}{074505}
  (\bibinfo{year}{2012}).

\bibitem{VovelleBergeron2000}
\bibinfo{author}{Bergeron, V.}, \bibinfo{author}{Bonn, D.},
  \bibinfo{author}{Martin, J.~Y.} \& \bibinfo{author}{Vovelle, L.}
\newblock \bibinfo{title}{Controlling droplet deposition with polymer
  additives}.
\newblock \emph{\bibinfo{journal}{Nature}} \textbf{\bibinfo{volume}{405}},
  \bibinfo{pages}{772--775} (\bibinfo{year}{2000}).

\bibitem{WangHao2015}
\bibinfo{author}{Hao, C.~L.} \emph{et~al.}
\newblock \bibinfo{title}{Superhydrophobic-like tunable droplet bouncing on
  slippery liquid interfaces}.
\newblock \emph{\bibinfo{journal}{Nat. Commun.}} \textbf{\bibinfo{volume}{6}},
  \bibinfo{pages}{7986} (\bibinfo{year}{2015}).

\bibitem{TakeharaThoroddsen2008}
\bibinfo{author}{Thoroddsen, S.}, \bibinfo{author}{Etoh, T.} \&
  \bibinfo{author}{Takehara, K.}
\newblock \bibinfo{title}{High-speed imaging of drops and bubbles}.
\newblock \emph{\bibinfo{journal}{Annu. Rev. Fluid Mech.}}
  \textbf{\bibinfo{volume}{40}}, \bibinfo{pages}{257--285}
  (\bibinfo{year}{2008}).

\bibitem{ButtDeng2012}
\bibinfo{author}{Deng, X.}, \bibinfo{author}{Mammen, L.},
  \bibinfo{author}{Butt, H.~J.} \& \bibinfo{author}{Vollmer, D.}
\newblock \bibinfo{title}{Candle soot as a template for a transparent robust
  superamphiphobic coating}.
\newblock \emph{\bibinfo{journal}{Science}} \textbf{\bibinfo{volume}{335}},
  \bibinfo{pages}{67--70} (\bibinfo{year}{2012}).

\bibitem{RobinVerho2012}
\bibinfo{author}{Verho, T.} \emph{et~al.}
\newblock \bibinfo{title}{Reversible switching between superhydrophobic states
  on a hierarchically structured surface}.
\newblock \emph{\bibinfo{journal}{Proc. Natl. Acad. Sci.}}
  \textbf{\bibinfo{volume}{109}}, \bibinfo{pages}{10210--10213}
  (\bibinfo{year}{2012}).

\bibitem{AizenbergMishchenko2010}
\bibinfo{author}{Mishchenko, L.} \emph{et~al.}
\newblock \bibinfo{title}{Design of ice-free nanostructured surfaces based on
  repulsion of impacting water droplets}.
\newblock \emph{\bibinfo{journal}{ACS Nano}} \textbf{\bibinfo{volume}{4}},
  \bibinfo{pages}{7699--7707} (\bibinfo{year}{2010}).

\bibitem{JiangJu2012}
\bibinfo{author}{Ju, J.} \emph{et~al.}
\newblock \bibinfo{title}{A multi-structural and multi-functional integrated
  fog collection system in cactus}.
\newblock \emph{\bibinfo{journal}{Nat. Commun.}} \textbf{\bibinfo{volume}{3}},
  \bibinfo{pages}{1247} (\bibinfo{year}{2012}).

\bibitem{WangChen2011}
\bibinfo{author}{Chen, X.} \emph{et~al.}
\newblock \bibinfo{title}{Nanograssed micropyramidal architectures for
  continuous dropwise condensation}.
\newblock \emph{\bibinfo{journal}{Adv. Funct. Mater.}}
  \textbf{\bibinfo{volume}{21}}, \bibinfo{pages}{4617--4623}
  (\bibinfo{year}{2011}).

\bibitem{ChenJonathan2009}
\bibinfo{author}{Boreyko, J.~B.} \& \bibinfo{author}{Chen, C.-H.}
\newblock \bibinfo{title}{Self-propelled dropwise condensate on
  superhydrophobic surfaces}.
\newblock \emph{\bibinfo{journal}{Phys. Rev. Lett.}}
  \textbf{\bibinfo{volume}{103}}, \bibinfo{pages}{184501}
  (\bibinfo{year}{2009}).

\bibitem{Stone2012}
\bibinfo{author}{Stone, H.~A.}
\newblock \bibinfo{title}{Ice-phobic surfaces that are wet}.
\newblock \emph{\bibinfo{journal}{ACS Nano}} \textbf{\bibinfo{volume}{6}},
  \bibinfo{pages}{6536--6540} (\bibinfo{year}{2012}).

\bibitem{PoulikakosMaitra2014}
\bibinfo{author}{Maitra, T.} \emph{et~al.}
\newblock \bibinfo{title}{Supercooled water drops impacting superhydrophobic
  textures}.
\newblock \emph{\bibinfo{journal}{Langmuir}} \textbf{\bibinfo{volume}{30}},
  \bibinfo{pages}{10855--10861} (\bibinfo{year}{2014}).

\bibitem{PoulikakosMaitra2014-1}
\bibinfo{author}{Maitra, T.} \emph{et~al.}
\newblock \bibinfo{title}{On the nanoengineering of superhydrophobic and
  impalement resistant surface textures below the freezing temperature}.
\newblock \emph{\bibinfo{journal}{Langmuir}} \textbf{\bibinfo{volume}{14}},
  \bibinfo{pages}{1106--1106} (\bibinfo{year}{2014}).

\bibitem{QuereRichard2002}
\bibinfo{author}{Richard, D.}, \bibinfo{author}{Clanet, C.} \&
  \bibinfo{author}{Qu{\'e}r{\'e}, D.}
\newblock \bibinfo{title}{Contact time of a bouncing drop}.
\newblock \emph{\bibinfo{journal}{Nature}} \textbf{\bibinfo{volume}{417}},
  \bibinfo{pages}{811} (\bibinfo{year}{2002}).

\bibitem{Marmur2004}
\bibinfo{author}{Marmur, A.}
\newblock \bibinfo{title}{The lotus effect: Superhydrophobicity and
  metastability}.
\newblock \emph{\bibinfo{journal}{Langmuir}} \textbf{\bibinfo{volume}{20}},
  \bibinfo{pages}{3517--3519} (\bibinfo{year}{2004}).

\bibitem{Zwliu2014}
\bibinfo{author}{Liu, Y.~H.} \emph{et~al.}
\newblock \bibinfo{title}{Pancake bouncing on superhydrophobic surfaces}.
\newblock \emph{\bibinfo{journal}{Nat. Phys.}} \textbf{\bibinfo{volume}{10}},
  \bibinfo{pages}{515--519} (\bibinfo{year}{2014}).

\bibitem{WangLiu2015}
\bibinfo{author}{Liu, Y.~H.}, \bibinfo{author}{Whyman, G.},
  \bibinfo{author}{Bormashenko, E.}, \bibinfo{author}{Hao, C.~L.} \&
  \bibinfo{author}{Wang, Z.~K.}
\newblock \bibinfo{title}{Controlling drop bouncing using surfaces with
  gradient features}.
\newblock \emph{\bibinfo{journal}{Appl. Phys. Lett.}}
  \textbf{\bibinfo{volume}{107}}, \bibinfo{pages}{051604}
  (\bibinfo{year}{2015}).

\bibitem{JuliaMoevius2014}
\bibinfo{author}{Moevius, L.}, \bibinfo{author}{Liu, Y.~H.},
  \bibinfo{author}{Wang, Z.~K.} \& \bibinfo{author}{Yeomans, J.~M.}
\newblock \bibinfo{title}{Pancake bouncing: simulations and theory and
  experimental verification}.
\newblock \emph{\bibinfo{journal}{Langmuir}} \textbf{\bibinfo{volume}{30}},
  \bibinfo{pages}{13021--13032} (\bibinfo{year}{2014}).

\bibitem{Rayleigh1879}
\bibinfo{author}{Rayleigh, L.}
\newblock \bibinfo{title}{On the capillary phenomena of jets}.
\newblock \emph{\bibinfo{journal}{Proc. R. Soc. Lond.}}
  \textbf{\bibinfo{volume}{29}}, \bibinfo{pages}{71--79}
  (\bibinfo{year}{1879}).

\bibitem{Wangchu2010}
\bibinfo{author}{Chu, K.~H.}, \bibinfo{author}{Xiao, R.} \&
  \bibinfo{author}{Wang, E.~N.}
\newblock \bibinfo{title}{Uni-directional liquid spreading on asymmetric
  nanostructured surfaces}.
\newblock \emph{\bibinfo{journal}{Nat. Mater.}} \textbf{\bibinfo{volume}{9}},
  \bibinfo{pages}{413--417} (\bibinfo{year}{2010}).

\bibitem{Chendaniel2001}
\bibinfo{author}{Daniel, S.}, \bibinfo{author}{Chaudhury, M.~K.} \&
  \bibinfo{author}{Chen, J.~C.}
\newblock \bibinfo{title}{Fast drop movements resulting from the phase change
  on a gradient surface}.
\newblock \emph{\bibinfo{journal}{Science}} \textbf{\bibinfo{volume}{291}},
  \bibinfo{pages}{633--636} (\bibinfo{year}{2001}).

\bibitem{Demirelmalvadkar2010}
\bibinfo{author}{Malvadkar, N.~A.}, \bibinfo{author}{Hancock, M.~J.},
  \bibinfo{author}{Sekeroglu, K.}, \bibinfo{author}{Dressick, W.~J.} \&
  \bibinfo{author}{Demirel, M.~C.}
\newblock \bibinfo{title}{An engineered anisotropic nanofilm with
  unidirectional wetting properties}.
\newblock \emph{\bibinfo{journal}{Nat. Mater.}} \textbf{\bibinfo{volume}{9}},
  \bibinfo{pages}{1023--1028} (\bibinfo{year}{2010}).

\bibitem{Demirelmalvadkar2014}
\bibinfo{author}{Liu, C.}, \bibinfo{author}{Ju, J.}, \bibinfo{author}{Zheng,
  Y.} \& \bibinfo{author}{Jiang, L.}
\newblock \bibinfo{title}{Asymmetric ratchet effect for directional transport
  of fog drops on static and dynamic butterfly wings}.
\newblock \emph{\bibinfo{journal}{ACS Nano}} \textbf{\bibinfo{volume}{8}},
  \bibinfo{pages}{1321--1329} (\bibinfo{year}{2014}).

\bibitem{WangMalouin2010}
\bibinfo{author}{Malouin, B.~A.}, \bibinfo{author}{Koratkar, N.~A.},
  \bibinfo{author}{Hirsa, A.~H.} \& \bibinfo{author}{Wang, Z.}
\newblock \bibinfo{title}{Directed rebounding of droplets by microscale surface
  roughness gradients}.
\newblock \emph{\bibinfo{journal}{Appl. Phys. Lett.}}
  \textbf{\bibinfo{volume}{94}}, \bibinfo{pages}{234103}
  (\bibinfo{year}{2010}).

\bibitem{StoneBird2009}
\bibinfo{author}{Bird, J.~C.}, \bibinfo{author}{Tsai, S. S.~H.} \&
  \bibinfo{author}{Stone, H.~A.}
\newblock \bibinfo{title}{Inclined to splash: triggering and inhibiting a
  splash with tangential velocity}.
\newblock \emph{\bibinfo{journal}{New J. Phys.}} \textbf{\bibinfo{volume}{11}},
  \bibinfo{pages}{063017} (\bibinfo{year}{2009}).

\bibitem{WangChen2005}
\bibinfo{author}{Chen, R.~H.} \& \bibinfo{author}{Wang, H.~W.}
\newblock \bibinfo{title}{Effects of tangential speed on low-normal-speed
  liquid drop impact on a non-wettable solid surface}.
\newblock \emph{\bibinfo{journal}{Exp. Fluids}}
  \textbf{\bibinfo{volume}{39}}, \bibinfo{pages}{754--760}
  (\bibinfo{year}{2005}).

\bibitem{ChenZhang2015}
\bibinfo{author}{Zhang, K.} \emph{et~al.}
\newblock \bibinfo{title}{Self-propelled droplet removal from hydrophobic
  fiber-based coalescers}.
\newblock \emph{\bibinfo{journal}{Phys. Rev. Lett.}}
  \textbf{\bibinfo{volume}{115}}, \bibinfo{pages}{074502}
  (\bibinfo{year}{2015}).

\bibitem{Vignes-AdlerLorenceau2009}
\bibinfo{author}{Lorenceau, E.}, \bibinfo{author}{Clanet, C.},
  \bibinfo{author}{Qu{\'e}r{\'e}, D.} \& \bibinfo{author}{Vignes-Adler, M.}
\newblock \bibinfo{title}{Off-centre impact on a horizontal fibre}.
\newblock \emph{\bibinfo{journal}{Eur. Phys. J-Spec Top.}}
  \textbf{\bibinfo{volume}{166}}, \bibinfo{pages}{3--6} (\bibinfo{year}{2009}).

\bibitem{VandewalleGilet2010}
\bibinfo{author}{Gilet, T.}, \bibinfo{author}{Terwagne, D.} \&
  \bibinfo{author}{Vandewalle, N.}
\newblock \bibinfo{title}{Droplets sliding on fibres}.
\newblock \emph{\bibinfo{journal}{Eur. Phys. J. E.}}
  \textbf{\bibinfo{volume}{31}}, \bibinfo{pages}{253--262}
  (\bibinfo{year}{2010}).

\bibitem{QuereClanet2004}
\bibinfo{author}{Clanet, C.}, \bibinfo{author}{B{\'e}guin, C.},
  \bibinfo{author}{Richard, D.} \& \bibinfo{author}{Qu{\'e}r{\'e}, D.}
\newblock \bibinfo{title}{Maximal deformation of an impacting drop}.
\newblock \emph{\bibinfo{journal}{J. Fluid Mech.}}
  \textbf{\bibinfo{volume}{517}}, \bibinfo{pages}{199--208}
  (\bibinfo{year}{2004}).

\bibitem{Lintaehun2010}
\bibinfo{author}{Lee, T.} \& \bibinfo{author}{Liu, L.}
\newblock \bibinfo{title}{Lattice boltzmann simulations of micron-scale drop
  impact on dry surface}.
\newblock \emph{\bibinfo{journal}{J. Comput. Phys.}}
  \textbf{\bibinfo{volume}{229}}, \bibinfo{pages}{8045--8063}
  (\bibinfo{year}{2010}).

\bibitem{TaehunKevin2013}
\bibinfo{author}{K., C.} \& \bibinfo{author}{Lee, T.}
\newblock \bibinfo{title}{Lattice boltzmanm simulations of forced wetting
  transitions of drops on superhydrophobic surface}.
\newblock \emph{\bibinfo{journal}{J. Comput. Phys.}}
  \textbf{\bibinfo{volume}{250}}, \bibinfo{pages}{601--615}
  (\bibinfo{year}{2013}).

\bibitem{Bonnbartolo2005}
\bibinfo{author}{Bartolo, D.}, \bibinfo{author}{Josserand, C.} \&
  \bibinfo{author}{Bonn, D.}
\newblock \bibinfo{title}{Retraction dynamics of aqueous drops upon impact on
  non-wetting surfaces}.
\newblock \emph{\bibinfo{journal}{J. Fluid Mech.}}
  \textbf{\bibinfo{volume}{545}}, \bibinfo{pages}{329--338}
  (\bibinfo{year}{2005}).

\bibitem{Savage2015}
\bibinfo{author}{Savage, N.}
\newblock \bibinfo{title}{Synthetic coatings: Super surfaces}.
\newblock \emph{\bibinfo{journal}{Nature}} \textbf{\bibinfo{volume}{519}},
  \bibinfo{pages}{S7--S9} (\bibinfo{year}{2015}).

\bibitem{BourouibaGilet2014}
\bibinfo{author}{Gilet, T.} \& \bibinfo{author}{Bourouiba, L.}
\newblock \bibinfo{title}{Rain-induced ejection of pathogens from leaves:
  Revisiting the hypothesis of splash-on-film using high-speed visualization}.
\newblock \emph{\bibinfo{journal}{Integr. Comp. Biol.}}
  \textbf{\bibinfo{volume}{54}}, \bibinfo{pages}{974--984}
  (\bibinfo{year}{2014}).

\end{thebibliography}
\end{document}